\documentclass{JHEP3} 

\usepackage{graphicx}

\preprint{Liverpool Preprint: LTH XXX}

\title{Searching for chiral logs in the static-light decay constant.}

\author{UKQCD Collaboration: C. McNeile, and C.~Michael \\
Theoretical Physics Division, Dept. of Mathematical Sciences, 
          University of Liverpool, Liverpool L69 3BX, UK 
}

\keywords{Lattice QCD}

\abstract{
  Using the clover fermion action in unquenched QCD with pion
  masses as low as 420 MeV, we look for evidence for chiral logs in
  the static-light decay constant.  There is some evidence for a
  chiral log term, if the original static theory of Eichten and Hill
  is used. However, the more precise data from the static action of
  the ALPHA collaboration do not show any evidence for non-linear
  dependence of the static-light decay constant on the light quark
  mass.  We make some comments on the connection between chiral
  perturbation theory for decay constants of the pion and static-light meson.}
\begin{document}

\section{Introduction}

The data from experiments such as BaBar and Belle is helping  to measure
the CKM matrix better  (and hopefully see a breakdown of the standard
model formalism). To extract information about the quarks, QCD must be
solved for various non-perturbative matrix elements. In particular,  the
ratio of the decay constants of the $B_s$ to $B$ mesons 
($\frac{f_{B_s}}{f_B}$) is a crucial QCD quantity  for the unitarity
checks of the CKM matrix. It will become more  important once $B_s$
mixing is directly measured at run II of  the Tevatron.
Ali~\cite{Ali:2003te} and Lubicz~\cite{Lubicz:2004nn} review the
dependence of the QCD matrix elements  $\frac{f_{B_s}}{f_B}$ on the
determination of $\mid V_{td} \mid $ and $\mid V_{ts} \mid $.

There used to a complaisant view (with perhaps a few
exceptions~\cite{Atwood:2001jr,Sharpe:1995qp}) that  the ratio  of
$\frac{f_{B_s}}{f_B}$ could  easily be computed  reliably from lattice
QCD, because systematic errors would be reduced in ratios of decay
constants.

The error on the ratio of the $\frac{f_{B_s}}{f_B}$ has recently been
increased, however, because the uncertainty due to the long
extrapolation in the quark mass was
underestimated~\cite{Kronfeld:2002ab}. For example,  the
JLQCD~\cite{Aoki:2003xb}  collaboration quote 
$f_{B_s}/f_{B_d}=1.13(3)(^{+13}_{-\ 2})$, where the first error is
statistical and the second error is that from the systematic
uncertainties. The dominant systematic uncertainty in JLQCD's result is
from the chiral extrapolation of $f_B$ to light quark mass. There has
also been work where the ratio of  the $B$ meson decay constant to the
pion decay constant is used to control the log
terms~\cite{Bernard:2002pc,Becirevic:2002mh}.

The problem is extrapolating the value of the $f_B$ decay constant from
the masses in lattice calculations to the physical point. In particular
heavy-light chiral perturbation theory predicts  a log term in the light
quark mass dependence of $f_B$. All previous lattice calculations, apart
from some preliminary evidence from the HPQCD
group~\cite{Wingate:2003gm,Gray:2004hd}, have only seen linear
dependence  of the heavy-light decay constant on the quark mass.

There have been a number of attempts to estimate the  error from
extrapolating down to the light  quark masses, using some physically
modified form of chiral perturbation 
theory~\cite{Aoki:2003xb,Kronfeld:2002ab,Bernard:2002pc,Becirevic:2002mh}.
These have been criticised by  Sanz-Cillero et
al.~\cite{Sanz-Cillero:2003fq} who  claim that the systematic
uncertainty due  to the chiral extrapolation may have been 
overestimated by JLQCD~\cite{Aoki:2003xb} and  Kronfeld and
Ryan~\cite{Kronfeld:2002ab}. Rather than blindly introducing the chiral
log term to lattice data when those data do not show any sign of a
departure from a linear behaviour, it would be better to resolve these
issues by  explicit looking for non-linear  dependence of the
static-light decay constant on the mass of the light quark and this is
the  goal of  our unquenched lattice QCD calculations.

The UKQCD collaboration have recently  finished a calculation that used 
sea quarks with masses around a third of the strange quark
mass~\cite{Allton:2004qq}.  This is lighter than the sea quark masses
used by the JLQCD calculation~\cite{Aoki:2003xb}.  As part of that study
 UKQCD claimed to see  some evidence for the chiral log term in the pion
decay  constant~\cite{Allton:2004qq}. As has been noted by many 
authors~\cite{Bernard:2002pc,Becirevic:2002mh} the chiral log structure 
of $f_\pi$  and $f_B$ is rather similar at one loop. Hence, a detection
of  chiral logs in $f_\pi$ is an indication that the parameters of the
unquenched calculation are close to where chiral logs may occur  in the
heavy-light decay constant. The value of the lightest pion in  this
calculation is  roughly 420 MeV~\cite{Allton:2004qq}. The different
treatments of the heavy-light chiral perturbation theory of Sanz-Cillero
et al.~\cite{Sanz-Cillero:2003fq} show that a deviation of linearity is
expected at these pion masses.

The improved staggered formalism has produced  gauge configurations with
pion masses as  light as  320 MeV~\cite{Aubin:2004wf}. Heavy-light
staggered chiral perturbation theory can produce non-intuitive
results~\cite{Aubin:2004xd}.
 It is valuable to perform cross checks on results  from improved
staggered calculations with another fermion formalism, irrespective of
any theoretical concerns about the  improved staggered 
formalism~\cite{Jansen:2003nt}. The huge computational costs of
unquenched calculations with Wilson and domain  wall fermions makes this
a tough goal~\cite{Jansen:2003nt,Kennedy:2004ae}.

\section{Numerical methods}

The basis of our calculation is unquenched  gauge configurations
generated with  the non-perturbatively improved clover action and the
Wilson gauge action. The lattice parameters are: volume $16^3 \; 32$, 
$\beta$ = 5.2, and  the  clover coefficient was the non-perturbative
value of 2.0171. We only use the same sea and valence quark mass. The
full details of the action and results on the hadron spectroscopy have 
been published~\cite{Allton:2004qq,Allton:2001sk}.

We use static quarks for the heavy mesons. The static formalism is the
ideal framework for investigating the log form. It has fewer parameters
in the effective Lagrangian (because the parameters due to  $1/M_Q$
terms are obviously absent). As noted by Arndt and
Lin~\cite{Arndt:2004bg}, finite size effects are reduced in the static
limit. Also the exploration of chiral logs in the light  quark sector by
CP-PACS~\cite{Namekawa:2004bi} and qq+q
collaboration~\cite{Farchioni:2004tv}, essentially used a relatively
coarse lattice spacing. It is very difficult to apply these techniques 
for heavy quark actions because of large $a M_Q$ errors.

UKQCD has already published~\cite{Green:2003zz} an extensive analysis of
the spectrum of static-light mesons and a paper on the mass of the 
bottom quark~\cite{McNeile:2004cb}. In this paper we look for chiral
logs in the  heavy-light decay constant.

As the aim of this work is to look for chiral logs in  the $f_B$ decay
constant that are a small effect, it is important to reduce the
statistical errors.  In our previous calculations we were already using
all-to-all and fuzzing techniques, so we needed a new method to reduce
the statistical errors. The number of available gauge configurations is
fixed. The ALPHA collaboration~\cite{DellaMorte:2003mn} have developed 
a new variant of the static formalism that reduces the  $1/a$ mass
renormalisation that is thought to be the reason for  the poor signal to
noise  ratio of static-light calculations.

 The static action is given by
\begin{equation}
S_h = a^4 \sum_x \overline{\psi}_h(x) D_0 \psi_h(x)
\end{equation}

\begin{equation}
 D_0 \psi_h(x) = \frac{1}{a} [ \psi_h(x)  - W^\dagger(x - a
 \hat{t})_{t} ] \psi_h(x - \hat{t})
\end{equation}
The original static action written down by Eichten and  Hill used
$W(x)_t$ = $U(x)_t$. The version of the ALPHA~\cite{DellaMorte:2003mn} 
static action that we investigated was:
\begin{equation}
V(x)_t = \frac{1}{c}
\sum_{j=1}^{3}
[ 
U(x)_j U(x + a \hat{j})_t  U(x + a \hat{t} )^\dagger_j
+
U(x - a \hat{j} )_j^\dagger U(x - a \hat{j})_t  U(x + a \hat{t} - a \hat{j})_j
]
\end{equation}
 where $c$ is a normalisation constant. This is their $A$ variant  of
the static-light action, here labelled by suffix A or called "fuzzed
static".

The $f_B$ decay constant is defined by the matrix element 
below:
 \begin{equation}
\langle 0 \mid A_\mu \mid B(p) \rangle = i p_\mu f_B
\end{equation}
The $f_B$ matrix element is extracted from 
the amplitudes in the two point correlator.
\begin{eqnarray}
C(t) & = &  \sum_{x} 
\langle 0 \mid 
A_4(x,t)
\Psi_B^\dagger(x,0) 
\mid 0 \rangle \\
     & \rightarrow & Z_L Z_{\Psi_B} \exp( - a{ \cal E } t)
\label{eq:fitModel}
\end{eqnarray}
 where  $\Psi_B$ is the interpolating operator for  static-light mesons
and ground state dominance is shown in equation~\ref{eq:fitModel}. The
$Z_L$ amplitude is related to the $f_B^{static}$ decay constant
 \begin{equation}
f_B^{static} = Z_L \sqrt{ \frac{2}{M_B} } Z_A^{static}
 \end{equation}
 We discuss the renormalisation factor $Z_A^{static}$ later.
We~\cite{Green:2003zz} used  all-to-all
propagators~\cite{Michael:1998sg} and  fuzzed sources  to get accurate
correlators. We fit a 5 exponential model to a 5 by 5 smearing matrix.
We discuss the fit strategy  in more detail in section~\ref{eq:results}.

\section{Improvement and perturbative matching}

To extract $f_B^{stat}$ we need the 
renormalisation 
and $O(a)$ improvement terms for 
the static-light axial current.
\begin{equation}
A^{stat}_0 = 
\overline{\psi} \gamma_k \gamma_5 
\psi_Q
\end{equation}
The improved static current is:
\begin{equation}
(A_I^{stat})_0 = A_{0}^{stat} 
+ a c_A^{stat}
\overline{\psi} \gamma_k \gamma_5 \frac{1}{2} 
(\overleftarrow{D}_k +\overleftarrow{D}_k^\dagger ) \psi_Q
\label{eq:impStatLight}
\end{equation}
in the ALPHA formalism~\cite{Kurth:2000ki}.
$\overleftarrow{D}_k$ are covariant derivatives acting on the light
quark fields ($\psi$). The improvement term in
equation~\ref{eq:impStatLight} was first introduced by 
Morningstar and 
Shigemitsu~\cite{Morningstar:1998ep}.

To get from the static theory on the lattice to the 
static theory in the continuum a renormalisation
factor is required.
\begin{equation}
(A_R^{stat})_0 = 
Z_A^{stat} (1 + b_A^{stat} a m_q ) 
(A_I^{stat})_0 
\end{equation}
where $m_q$ is the light quark mass.

 The improvement coefficients $b_A^{stat}$ and  $c_A^{stat}$ have been
computed to one loop in  perturbation
theory~\cite{Morningstar:1998ep,Kurth:2000ki}.
 \begin{equation}
b_A^{EH ; stat} = \frac{1}{2} - 0.056 g^2
 \end{equation}
We use the tree level value of $1/2$ for $b_A^{A ; stat}$, 
because the one loop calculation has not yet been 
completed.

\begin{equation}
c_A^{EH ; stat} = -\frac{1}{4 \pi} 1.0 g^2
\end{equation}

\begin{equation}
c_A^{A ; stat} = -0.1164 g^2
\end{equation}

UKQCD~\cite{Bowler:2000xw} 
have recently written down the $Z$ factors
in the static limit. These were obtained from
Kurth and Sommer's calculation~\cite{Kurth:2000ki}.
This was an update on the
Borrelli and Pittori~\cite{Borrelli:1992fy} calculation.


\begin{equation}
Z_A^{stat} = 1.0 + ( \frac{\log(a\mu) }{4 \pi^2} - 0.137 ) g^2
\end{equation}

As noted by Hernandez and Hill~\cite{Hernandez:1994bx}, there is no
effect on the value of $Z_A$ from tadpole improvement if the standard
exponential fit model is used to extract the amplitude from the
correlators. In principle the improvement term could be 
tadpole improved.

As we are interested in the chiral logs in the 
leading order heavy-light chiral Lagrangian, we 
don't include a matching factor from the continuum
static theory to QCD~\cite{Hernandez:1994bx,Duncan:1995uq}.
The MILC collaboration~\cite{Bernard:2002pc} 
have recently discussed the appropriate scale 
for $Z_A^{stat}$
using the Lepage-Mackenzie scale setting procedure~\cite{Lepage:1993xa}.

The perturbative expression for $Z_A$ for the ALPHA static action is not
yet available. Experience with  other ``fuzzed'' fermion actions
suggests that the  value of $Z_A$ will be closer to one than for the 
static heavy quark action of  Eichten and Hill~\cite{Bernard:2002pc,Bernard:1999kc}. 
The ``smearing'' out of
the gauge links can be thought of as averaging over the fields. This
reduces  the perturbative corrections. Alternatively, the ``smearing''
can be thought of as  smoothing the potential in which the light quark
moves, so reducing the wave-function at  the origin, related to $f_B$.

It was found by the MILC 
collaboration~\cite{Bernard:2002pc,Bernard:1999kc},  for a smeared
version of the clover action, that too much smearing of the  fields can
drastically change the decay constant. For the  static action introduced
by the ALPHA collaboration, only one level of smearing is used. So the
heavy quark is restricted to  lie within $\pm a$ of the origin. At our
lattice spacing, the spatial extent of the light quark in a heavy-light
meson  is typically $3a$ (for example the node in the excited wave
function is at that distance~\cite{Michael:1998sg}). Thus we expect 
this smearing of the heavy quark to retain the same  qualitative
features,  but it should affect the renormalisation factor for the
action. This picture is consistent with the reported comparison of
$f_{B_s}$ computed in quenched QCD at $\beta = 6.2$ by Abada et
al.~\cite{Abada:2003un}, using both the Eichten and Hill static action
and the static action introduced by ALPHA. However, our results  using
the Eichten-Hill static quark action should be  less contaminated by
excited states due to the use of all-to-all  propagators and variational
smearing techniques.

\section{Chiral perturbation theory} \label{eq:chiralPert}

The static limit is the ideal place to
study the chiral logs predicted by heavy-light
chiral perturbation theory.
\begin{equation}
\Phi_{f_{B_d}}
\equiv f_{B_d} \sqrt{M_{B_d}}
\end{equation}

The one loop correction to the static-light decay constant for 2 flavours
of degenerate fermions with lightest pseudoscalar mass labelled $m_{\pi}$
is~\cite{Aoki:2003xb,Grinstein:1992qt,Goity:1992tp}
\begin{equation}
\frac
{\Phi_{f_{B_d}}}
{\Phi_{f_{B_d}^0}}
= 1 - \frac{3(1 + 3 g^2) }{4} \frac{ m_\pi^2 } { (4 \pi f )^2 }
\log( \frac{ m_\pi^2  } { \mu^2} )
\end{equation}
 where $g$ is  the $B^{\star}B\pi$ coupling. There have been a number of
 calculations~\cite{deDivitiis:1998kj,Abada:2002xe,Abada:2003un} of the
coupling $g$ using quenched  lattice QCD. We use the nominal value of
$g^2 = 0.35$ in this analysis.

In order to compare with our lattice results, we use the expressions
derived by  Sanz-Cillero et al.~\cite{Sanz-Cillero:2003fq} with a fixed
cut-off $\Lambda$  since this emphasises the region which is reliable in
chiral perturbation theory.  As well as the one loop correction from the
lowest  order chiral Lagrangian, there will be a term from a higher
contribution to the chiral Lagrangian, namely $\alpha_1 m^2_{\pi}$.
Thus there  are three free parameters: $\Lambda$, $\alpha_1$ and
$\Phi_0$. In order to  establish reasonable values for these parameters,
we fit this chiral expression  at our two heavier quark masses, for each
of the two values of $\Lambda=0.4$  and 1.0 GeV, so determining a range
of predictions for lighter quark masses.  

\section{Results} \label{eq:results}

In table~\ref{tab:bareRESULTS} we present our results  for the
static-light decay constant as a function of sea $\kappa$ value. These
values come from fitting our $5\times5$ matrix or correlators  over the
$t$-range 4-15 with a 5 exponential expression. This sophisticated
treatment  is optimal to deal with the excited state spectrum and to
extract cleanly the  ground state contribution needed to determine
$Z_L$. The $\chi^2/\mathrm{d.o.f.}$ is acceptable (indeed for
$\kappa=0.1350$, we also get an acceptable $\chi^2$ for the naive static
case for $t$-range 3-15, and that result is  tabulated).

 Our results  show clearly that the fuzzed static source advocated by
the  ALPHA collaboration does give a much smaller  statistical error.

\TABLE{
\begin{tabular}{|c|c|cc|cc|cc|}
Name & No.& $r_0 m_{PS}$ & $\kappa$ & 
$Z_L^0$ &  $r_0^{3/2}Z_L$ &
$\hat{Z}_L^0$ &  $r_0^{3/2}\hat{Z}_L$ \\ \hline
DF3 & 160 & 1.93(3) & 0.1350 & 
$0.304_{-10}^{+15}$ & 2.87(11) & $0.215_{-4}^{+4}$ & 1.99(3)\\
DF4 & 119 & 1.48(3) & 0.1355 & 
$0.282_{-15}^{+14}$ & 2.89(15) & $0.179_{-6}^{+6}$ & 1.79(6)\\
DF6 & 139 & 1.06(3) & 0.1358 & 
$0.202_{-17}^{+22}$ & 2.24(21) & $0.159_{-5}^{+6}$ & 1.71(6) \\
  \end{tabular}
  \caption{
 Decay constants (excluding any factors of $Z_A$) for the naive static
source ($Z$) and the smeared source ($\hat{Z}$) for each data set. The
uncorrected  lattice value is $Z^0_L$ (in lattice units), while the
improved decay constant is $Z_L$ (in units of $r_0$). 
}
\label{tab:bareRESULTS}
}

We compare in fig.~\ref{fig:heavyLog} the chiral prediction discussed
above with our lattice data for different quark masses, where the
heaviest quark mass  corresponds approximately to the strange mass. 
This chiral prediction is arranged to have a slope such that it goes
through the  values at our two larger quark masses for the fuzzed-static
case. As can be seen, the curvature expected from the chiral logarithm
should  have set in at our lightest quark mass. The statistical errors
are not  sufficiently small to verify this accurately, although there is
intriguing  sign of a substantial curvature for the naive static case.
Since this curvature  is not reproduced by the relatively more accurate
data from the fuzzed static case, we  caution that it may be a
statistical fluctuation.

 If we use the chiral model as a guide to the possible extrapolation to
lighter quark masses, as discussed above and shown in
fig.~\ref{fig:heavyLog}, we would obtain  $\frac{Z_L(B_s)}{Z_L(B)} $
from 1.31 to 1.46 as $\Lambda$ is varied  from 0.4 to 1.0 GeV. Thus we
have a systematic error of $\pm 5 \% $  arising from the chiral
extrapolation. We also have a  statistical error on the slope  which is 
larger at around $\pm 35 \%$. So we obtain  $\frac{f_{B_s}}{f_B} = 1.38
\pm 0.13 \pm 0.08$.
 This is to be compared with  Kronfeld's~\cite{Kronfeld:2002ab}  best
guess from JLQCD and HPQCD results of  $\frac{f_{B_s}}{f_B} = 1.25 \pm
0.10$.

\FIGURE{
\includegraphics[scale=0.6]{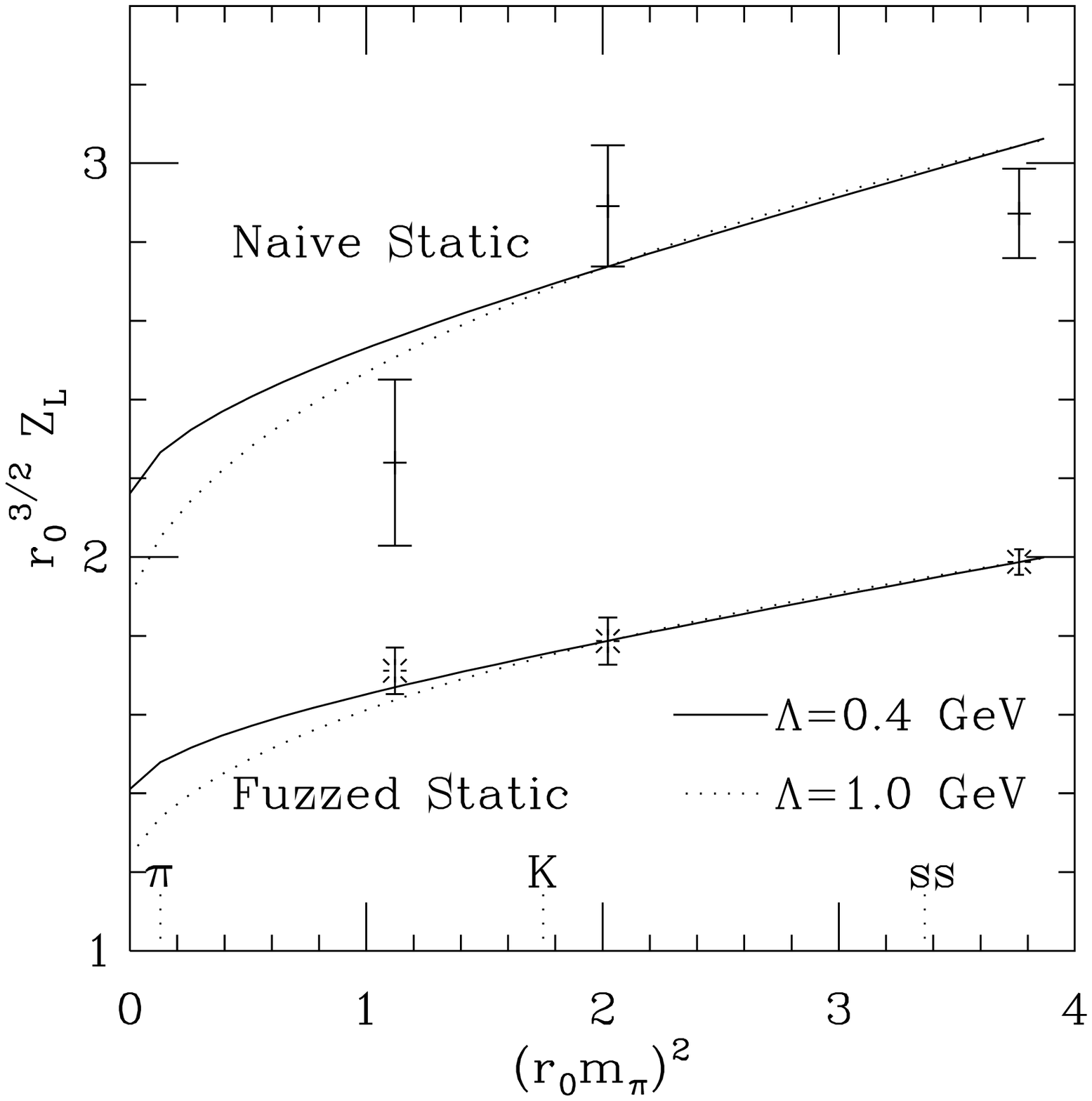}
\caption {
Static-light decay constant as a function of quark mass.
\label{fig:heavyLog}
}
}

Our primary aim is explore the light quark mass dependence of the
heavy-light decay constant, rather than to produce new values for the
decay constant. As a cross-check on our work, we compute the  $f_{B_s}$
decay constant for static (EH) $b$ quarks. The $DF3$ data set has sea quarks
with masses that are close to the mass of the strange quark. As
discussed in the previous section we use $Z_A$ = 0.68. Using $r_0$ =
0.55 (0.05) fm, we get $f_{B_s}^{stat}$ = 256  MeV with errors: 4\%
statistical, 14\%  from the scale $r_0$ and  10\% from the uncertainties
in $Z_A$. This value is reasonably consistent with other determinations of
$f_{B_s}^{stat}$. ALPHA obtain $f_{B_s}$ = 225 MeV in the continuum
limit of  quenched QCD using $r_0$ = 0.5 fm with a non-perturbative
renormalisation scheme. 
 Duncan et al.~\cite{Duncan:1995uq} obtain $f_{B_s}^{stat}$ = 304 MeV at
$a^{-1}$ = 1.78 GeV in quenched QCD using the Wilson action for the 
light quarks.

In UKQCD's work on chiral logs in the pion decay constant, there was a
concern about finite volume effects~\cite{Allton:2004qq}. At
$\kappa_{sea}$ = 0.1358 it was argued from chiral perturbation theory
that the finite volume effects were of the order of 8\% in $f_\pi$. A
similar order of magnitude effect was also estimated by Colangelo and
Haefeli~\cite{Colangelo:2004xr}. The volume of the lightest data set
$DF6$ is $(1.5 \mathrm{fm})^3$. Recently Arndt and
Lin~\cite{Arndt:2004bg,Goity:1990jb} have studied the  effect of the
finite volume on the ratio of heavy light decay  constants and bag
parameters. For a pion mass of around 400 MeV in a box of size  $(1.6
\mathrm{fm})^3$, they obtain a  finite volume effect in the ratio
$\frac{f_{B_s}}{f_B}$ of 0.006,  suggesting that the finite size
effects  are small. However, the next to leading order estimate  of
finite  size effects in $f_\pi$ was significant~\cite{Colangelo:2004xr}.
 Unfortunately,  Colangelo and Haefeli~\cite{Colangelo:2004xr} claim
that there is not enough information to make a similar estimate for 
$f_B$.

Because of computational limitations we are forced to work
at a fixed lattice spacing. 
It is difficult to estimate the uncertainty due to not doing a
continuum extrapolation. There are now variants 
of chiral perturbation theory that include the effects of the 
leading lattice spacing errors (see Bar~\cite{Bar:2004xp} 
for a review). Aubin and Bernard~\cite{Aubin:2004xd} 
have developed a formalism
for heavy-light chiral perturbation theory,
at non-zero lattice spacing,
with staggered
fermions as the light quarks.

\section{Conclusions} \label{eq:conc}

We have looked for the effects of the chiral log 
in the heavy-light decay constant using the lightest
unquenched clover sea quarks used to date. Unfortunately,
even with pions as light as 420 MeV, we do not see
any compelling evidence for the chiral log in 
the static-light decay constant.

We have seen that the quark mass dependence of the static-light decay
constant can be different to that of the pion decay constant. Both
quantities can in principle have a different volume dependence and
different higher order terms  in the chiral Lagrangian.  The similarity
of the one loop expressions for  $f_\pi$  and $f_B$ seems to be of 
limited value~\cite{Bernard:2002pc,Becirevic:2002mh} in guiding
extrapolation.

Although we have been unable to find unambiguous evidence of the chiral
logarithm,  the approach we have tried to take here is  potentially a
good way to study the  light quark mass dependence of the heavy-light
decay constant. The improvements in the numerical techniques for
static-light calculations means that quite precise results are
achievable. With current computer  resources, the only hope of working
with lighter sea masses with Wilson like quarks is to work at relatively
coarser lattice spacings. Using the  static limit for the heavy quarks
is  more controlled. The use of heavy-light perturbation theory at
nonzero lattice spacing would also be required.


\providecommand{\href}[2]{#2}\begingroup\raggedright\endgroup

%
\end{document}